\documentclass[a4paper,12pt]{article}
\usepackage{epsfig}
\usepackage{amssymb}
\begin{document}
\thispagestyle{empty}

\newcommand{\etal}  {{\it{et al.}}}  
\def\Journal#1#2#3#4{{#1} {\bf #2}, #3 (#4)}
\def\PRD{Phys.\ Rev.\ D}
\def\NIMA{Nucl.\ Instrum.\ Methods A}
\def\PRL{Phys.\ Rev.\ Lett.\ }
\def\PLB{Phys.\ Lett.\ B}
\def\EPJ{Eur.\ Phys.\ J}
\def\IEEETNS{IEEE Trans.\ Nucl.\ Sci.\ }
\def\CPCD{Comput.\ Phys.\ Commun.\ }


\bigskip

{\Large\bf
\begin{center}
Scalar boson stars: (thermo)dynamics and gravitational equilibria
\end{center}
}
\vspace{0.1 cm}

\begin{center}
{ G.A. Kozlov  }
\end{center}
\begin{center}
\noindent
 { Bogolyubov Laboratory of Theoretical Physics\\
 Joint Institute for Nuclear Research,\\
 Joliot Curie st., 6, Dubna, Moscow region, 141980 Russia  }
\end{center}
\vspace{0.1 cm}

 \begin{abstract}
 \noindent
 {We analyse the (thermo)dynamics of the scalar degrees of freedom  in the scalar boson stars through the dark matter density under extreme conditions.
The boson stars are studied in terms of a dark scalar sector in such a way this sector couples to the standard model Higgs boson doublet plus gravity.
The stability of the BS is investigated at the level of the interaction between the scalars with the scale invariance breaking triggered by the electroweak symmetry breaking plus gravity. 
Analytic methods have been applied to an effective version of the theory, the scalar "tower" approximation, which should preserve an exact scale invariance.
The production of the scalar dark matter and its decay have been discussed.
}


\end {abstract}




\bigskip

\section{Introduction}
Cosmological searches for dark sectors in terms of the hidden fields interacting feebly with the standard model (SM), still mainly focused on identifying signatures of the mediator fields and the couplings. In this aspect, the existence of dark matter (DM), in particular the bound states composed of the scalar DM, remains to be investigated both theoretically and experimentally using also the wide set of data from the collider and the fixed target experiments.    
The dark scalar fields could have the Bose-Einstein condensation (BEC) and form massive objects with large enough size as the scalar boson stars (BS) [1,2].
The BS as the cosmological object has the great interest in the sense to be explored in terms of  the observables and to be advocated and supported by the appropriate theoretical models (see, e.g., the refs. in [3,4]).
The origin of the BS, the mechanism of its formation and its lifetime are related with the scale invariance (SI) breaking, otherwise the SI stuff cannot have a mass unless that mass is zero. 
The highly successful models of the BS  still have the missing ingredients: the dark scalar fields and  their interactions with the SM Higgs bosons bounded by the self-interactions and the gravitational forces. During the formation of the boson stars, all the scales may depend on the properties of scaling dynamics associated with the new (dark) fields. 
In the SM, there is no way to understand the SI itself and to explore the properties of the BS.
The formation of the BS as the bound state of the cosmologically selected scalar fields, e.g., the Higgs bosons, the scalar dark matter (SDM), the dilaton-like particles, is related with the equilibrium relevant to gravitational forces and the potential interactions between these scalar fields. 
The SDM particles may appear either  as the decay product of other heavier particles in the dark sector, or due to fermionic DM $\chi$ annihilation into two SDM particles via exchanging $\chi$ in the $t$-channel.
An investigation of the SDM may be either through the relic density, the stable candidate to the DM, or within the observation of the cosmic rays due to decays of the SDM particles.
The starting point in the study of the BS is the gravity plus the Higgs potential  and the corrections to this potential accompanied by the cosmological dynamics of new scalar dark fields  in the approximate conformal symmetry breaking sector.
The scalar boson stars are just the giant scalar objects each of which  may consist of the huge number $N\sim 10^{(large\,number)}$ scalar particles held together  by the (long) gravitational attraction forces against the repulsion due to self-interaction of the SDM setting the scale for the breaking of scale invariance. 
The estimation of $N$ may be given if the critical temperature, the masses of the SDM and the BS $M^{\star}$  to be included in the calculations (see Sections 2 and 3).
The (mass) energy density of the BS is more than $10^{-10}$ orders of magnitude less than the energy density $\rho_{DM}$ of the local DM  in the vicinity of the solar system [4,5].
The central densities of the BS may exceed few times the ordinary scalar densities. In the interior of the BS the scalar particles distribute almost uniformly as a density $\rho_{DM}$.
One of the primary physics instruments to the BS studying  is to put the scalar fields under extreme physical conditions and to look how the properties will be changed with temperature $T$ and the mass (energy) density $\rho_{BS}$ of the BS.

When the BS is heated, it will experience the order-disorder (phase) transition at the critical temperature $T_c$ where the underlying hidden (conformal) symmetry becomes manifest. This phenomenon is rather universal in statistical physics, the particle physics  and in cosmology since the disordered phase with manifest symmetry  has lower free energy  at the sufficiently  high temperatures. In case of the BS, it is probable to realise the disordered scalar states if the star is highly excited since the number of such scalar states is far more greater than the ordered scalar states.  The spontaneous broken conformal symmetry will be restored as the BS is heated above certain $T_c$. When the symmetry breaking just starts, the BS would barely know which "direction" in mass and $T$ it should develop the BEC. This may cause the fluctuation of the condensate related to the particle density. 

One of the messengers between the SM and the DM belongs to an extra field sector with the light phantom $U^{\prime} (1)$ gauge bosons, the co-called dark photons (DP) (see, e.g., [6,7]). This new sector is hidden, the SM particles are not charged under  $U^{\prime} (1)$. The DP itself is the non-primary field operator because it can be defined through the derivative  of the SDM field [8]. The DP may play an essential role in the production of the SDM and its decay.
Both the mass of the DP and the kinetic strength responsible for the mixing between the DP with the ordinary photon are free parameters as well as the branching fraction $BR$ of the DP decay to the SM particles and to the invisible sector. The values of these  parameters and the $BR$ have to be estimated and determined phenomenologically, by comparing with the observations, and not theoretically. 

The plan of the paper is as follows. The Section 2 is devoted to the formation and the stability of the BS depending on the SDM density. The condensation of the scalars is considered in Sec. 3. The Sec. 4 sheds the light to the estimation of the observables.
In Sec.5 we develop the model with the minimal potential of the fields interactions plus gravity. The production of the SDM and the decay is presented in Sec. 6.
The paper is concluded in Sec. 7.

  \section{BS formation and the stationary point}
At temperatures $T=\beta^{-1}$ that are far away from that of the $T_c  > T$, close to the temperature of the conformal sector, the BS is in the statistical equilibrium where there are almost no interactions between the particles in the interior of the star. In this case, the partition function for $N$ scalar states  is $Z_{N} = Sp\, e^{-H \beta}$, 
where $H$ is the Hamiltonian for $N$ individual energies of each of particles, $H = \sum_{1\leq j\leq N} H(j)$.
For the system of $S_f (x)$ scalar fields which are regular functions in $f$-representation, one has the equation $H(j)\,S_{f}(x_{j}) = F(f)\,S_{f}(x_{j})$, where $H = \sum_{f}\,F(f)\,b^{+}_{f}\,b_{f} = \sum_{f}\,F(f)\,n_{f}$
is given in terms of the creation and the annihilation operators $b^{+}_{f}$ and $b_{f}$, respectively; $n_{f}$ is the occupation  number; $F(f) = E(f) - \mu\,Q (f)$ with $E(f)$ being the energy, $\mu$ is the chemical potential and 
$Q(f)$ is the conserved charge.
Since the BS has the thermal and the particle interactions with the reservoir, $Q$ is the operator $\hat N_{f}$ of the scalar particles of the type $f$ with the mean value 
$Tr \{\hat n\,\hat N_{f}\} = \rho_{f}\,\Omega_{BS}$ in the volume $\Omega_{BS}$ of the BS, where  $\hat n$ is the statistical operator,  $\rho_{f}$ is the density of the $f$-scalar boson.
The thermal equilibrium and the balance between the gravitational forces and the field  interactions  supported by large number $n_f$ may lead to the BEC and the formation of the BS (FBS).
The latter manifests itself through the energy (mass) density  $\bar\rho = \rho_{BS}/\rho_{DM}$ in the vicinity of the solar system, where $\rho_{BS}$ is the  density of the BS defined by its mass $M^{\star}$ and the number density $n^{\star}$, $\rho_{BS} = M^{\star} n^{\star}$. The DM density is $\rho_{DM} = \rho_{BS} + \mu_{s} n_s (1 +\delta_h)$, where $\mu_s$ and $n_s$ are the SDM mass and the number density, respectively, $\delta_h < 10^{-14}$ is the correction due to contribution from the SM Higgs boson $h$ to FBS.
The condition $\sum_{f} n_{f} = N$ can cause the difficulty to the factorisation  of the sum 
\begin{equation}
\label{e4}
Z_{N} = \sum_{{... n_{f} ... },{\sum_{f} n_{f} = N}} e^{-\beta\,\sum_{f} F({f})\,n_{f}}.
\end{equation}
It is also stressed by the fact that in the macroscopic system the particles may be created themselves and they can decay as well. Since all the  operators $... n_{f} ...$ in (\ref{e4}) commute to each other, they may be related with the observables. In order to calculate (\ref{e4}) at fixed $N$, one can use the asymptotic expressions for very large $N$. 
We can treat it as an effective description of a more complete model in which the BS is stabilised  at some background value of $\bar\rho$.    
To that end, as a concrete example, we introduce the probability of the FBS in the power series 
\begin{equation}
\label{e3}
P(\bar\rho) = \sum_{N =1}^{\infty} Z_{N}\,\bar\rho^{N}
\end{equation}
with
$$P(\bar\rho)  =   \sum_{...n_f ...}\bar\rho^{\sum_{f}n_{f}} e^{-\beta\sum_{f} F(f)n_{f}} =  \prod_{f} \left [\sum_{0\leq n <\infty}\bar\rho ^{n} e^{-F(f)n\beta}\right ] = \prod_f \frac{1}{1- \bar\rho \,e^{-F(f)\beta}}, $$
where $F(f)\geq  0$.
The chemical potential $\mu$ in $F(f)$ can vary over BS, but there is $\mu = const$ in $\Omega_{BS}$ if $\bar\rho\rightarrow 1$ and $\langle E(f)/\mu\rangle  \simeq \langle Q(f)\rangle\sim  \langle q\rangle\Omega_{BS}$ with $\langle q\rangle$ being the average density of $Q$ in the volume $\Omega_{BS}$.

Let us consider (\ref{e3})  in terms of the SDM density
\begin{equation}
\label{e6}
\frac{P(\bar\rho)}{\bar\rho^{N}} = \sum_{N^{\prime} =  0}^{\infty} \frac{Z_{N^{\prime}}\,\bar\rho^{N^{\prime}}}{\bar\rho^{N}}
\end{equation}
on the real axis $0 < \bar\rho < r_c$, where $r_c \geq 1$ is the convergence radius of (\ref{e3}).
 Because $Z_{N^{\prime}} > 0$, the function (\ref{e6}) has the only one minimum on $(0,r_c)$ 
$$\frac{d^{2}\left [P(\bar\rho)\bar\rho^{-N}\right ]} {d\bar\rho^{2}}= \sum_{N^{\prime} = 0}^{\infty} (N^{\prime} - N)(N^{\prime} - N -1)\,Z_{N^{\prime}}\bar\rho^{N^{\prime} - N -2} > 0. $$
The probability (\ref{e6}) tends to infinity when $\bar\rho\rightarrow 0$ and when $\bar\rho\rightarrow r_c$ at high enough $T$. Then, there is the point $\bar\rho = \bar\rho_{0}$ at which  (\ref{e6}) has a single minimum in the interval $(0, r_c)$, i.e. 
\begin{equation}
\label{e7}
  \frac{d}{d\bar\rho}\left [P(\bar\rho)\,\bar\rho^{-N}\right ]_{\vert_{\bar\rho =\bar\rho_{0}}} = \sum_{N^{\prime} = 0}^{\infty} Z_{N^{\prime}}\, (N^{\prime} - N)\,\bar\rho^{N^{\prime} - N -1}_{\vert_{\bar\rho = \bar\rho_{0}}} = 0. 
\end{equation}
If one goes through the point $\bar\rho = \bar\rho_0$ alone the vertical axis, the ratio (\ref{e6}) has a maximum at $\bar\rho_{0}$. 
The relative energy density $\bar\rho = \bar\rho_0$ is the stationary stable value when the BS is formed.

In the world of macroscopic objects, e.g., the BS, the spectrum of "quasi-momenta" $f$ will have the almost  continuous character in the limit $\Omega_{BS}\rightarrow\infty$. The particle number $\Delta N$ of different $\Delta f$  is proportional to 
\begin{equation}
\label{e8}
(\Delta N/\Delta f) = a\cdot \Omega_{BS},
\end{equation}
where $a$ is the positive constant.

Considering $P(\bar\rho)$ in the form $ P(\bar\rho) = \exp\left\{ -\sum_{f}\ln \left [ 1-\bar\rho\,e^{-F(f)\,\beta}\right ]\right\}, $
and taking into account (\ref{e8}) for large enough $N$,  one can get the asymptotic relation
$$\sum_{f}\ln \left [ 1-\bar\rho\,e^{-F(f)\,\beta}\right ] = N K_{DM} (\bar\rho),$$
where
\begin{equation}
\label{e9}
K_{DM}(\bar\rho) = a \cdot\nu \int \ln \left [ 1-\bar\rho\,e^{-F(f)\,\beta}\right ] df
\end{equation}
with $\nu = (\Omega_{BS}/N)$  = const
is the thermo-statistical contribution of the DM to the interaction potential where the integration in (\ref{e9}) is proceed over the momenta of the particles, $df\rightarrow d^{3}\vec p/(2\pi)^3.$
In order to scan the SDM density for the FBS we use the function

\begin{equation}
\label{e10}
 P(\bar\rho)\,\bar\rho^{-N} = \left [\bar\rho^{-1}\,e^{-K_{DM} (\bar\rho)}\right ]^{N}.
\end{equation}
The calculation of (\ref{e10}) may assist to find the condition for the formation of the BS and its stabilisation. 
To that end, we consider the circle $C$ in $\Omega_{BS}$ with the radius $r = \bar\rho_{0}$ where the origin is located at zero. 
We have
\begin{equation}
\label{e11}
 Z_{N} = \frac{1}{2 \pi i}\int_{C} \frac{P(\bar\rho)}{\bar\rho^{N+1}}\,d\bar\rho ,
\end{equation}
where doing the replacement $\bar\rho\rightarrow r e^{i\varphi}$, the maximum of the function under an integration in (\ref{e11}) is expected at $\varphi = 0$ 
taking into account the number of particles $N$ in the exponential function in (\ref{e10}). Hence, the maximal probability of the FBS in the asymptotic calculation  requires to know the behaviour  of the function under the integration in (\ref{e11}) at $\varphi = 0$.  
Note that 
$$\frac{\partial}{\partial\varphi}\left [ \frac{P(r\,e^{i\,\varphi})}{r^{N}\,e^{i\,N\,\varphi}}\right ]_{\vert_{\varphi = 0}} = 0 $$
is in force, because of the minimum condition (\ref{e7}). In the vicinity of $\varphi = 0$ the function under an integration has the form
\begin{equation}
\label{e12}
 \sim {\left [\frac{e^{- K_{DM} (r)}}{r}\right ]}^{N} e^{-N\alpha\,\varphi\,  +...},
\end{equation}
where $\alpha = K^{\prime}_{DM} (r) > 0$. We see from (\ref{e12}) that the main contribution to (\ref{e11}) is given by the values $\varphi\sim (\alpha N)^{-1}$ at $N\rightarrow \infty$. One can find asymptotically
 $$ Z_{N} (r) \simeq \frac{1}{2\pi} \int_{0}^{\infty} \frac{e^{- N K_{DM} (r)}}{r^{N}} e^{-N\alpha\varphi} d\varphi = \frac{e^{-N K_{DM} (r)}}{2\pi\alpha N r^{N}},$$
from which we obtain
\begin{equation}
 \label{e14}
 \ln Z_{N} (r)\simeq - \left \{\sum_{f}\,\ln \left [1-r\,e^{-F(f)\beta}\right ] + N \ln  r + \ln (2 \,\pi\,\alpha \, N)\right \}.
\end{equation}
Keeping all the terms in (\ref{e14}), we find asymptotically
\begin{equation}
 \label{e15}
\sum_{f} \bar n_{f} = \sum_{f} \frac{1}{\bar\rho_{0}^{-1}\, e^{F(f)\beta} -1} = N
\end{equation}
from which the $\bar\rho_{0}$ can be found. Using (\ref{e15}) one can calculate the sum of the scalar quantum states in the BS up to the condition of the BS stability at fixed $\bar\rho_{0}$. 
When $\bar\rho_{0} \rightarrow 1$, the large number $N$ is expected at high temperatures and when the relativistic scalar particles are light, that is important to the formation of the BS at early stage. There is the full condensate at very low $T$ and finite $\bar\rho_0$.
On the other hand, (\ref{e15}) will have the essential changes in the result if $T\rightarrow \infty$ when $\bar\rho_0\rightarrow 0$ (the phase transition). It can be induced by a strong magnetic field $B^{scalar}$ in the vacuum determined by the SDM mass $\mu_s$ at the EW scale, $B^{scalar}\sim \mu^{2}_s/e\sim O(10^{20}$ T), that might have been contributed to the formation of the Universe at its early stage. It may be happed that above $B^{scalar}$ the scale symmetry should be restored. 
The strong magnetic fields can modify the stability condition of the BS since they might have the cosmological meaning relevant to the EW phase transitions in the early Universe.
 These magnetic fields should also exist in the late Universe when the latter is under the influence of the magnetised black holes (see [9] and the refs. therein).

\section{The condensation and the stability condition for the BS }
The SDM particles can no get into chemical and thermal equilibrium during travelling in the early Universe before they take part in the FBS if these particles were produced in the decays of other heavier particles in the dark sector. 
As the temperature of the dark sector lowers below the scale $\sim \Lambda$ at which the SI is breaking down, the dark scalar particles can act themselves as the scalar degrees of freedom in the  classical form in the late Universe.  The particles are expected to lock up into the scalar massive objects, the BS.
Let us apply the results obtained in the Sec. 2 to the case of the non-relativistic model with  $N$ scalar particles, where these particles are in the volume $\Omega_{BS}$ as a cube with the side of the length $L \sim \Omega_{BS}^{1/3}$. We assume the wave function in the form 
$ \phi_{p} (q) = \Omega_{BS}^{-1/2}\,e^{i\,q\,p}$, where $p^{\alpha} = (2\,\pi/L)\,n^{\alpha}$, $\alpha = 1,2,3$; $n^{\alpha} = 0, \pm 1, ...$. The energy is $E_{p} = {\vert \vec p\vert}^{2}/(2\,\mu_{s})$ with the momentum $p$  (in the units with the Planck constant $h =1$).
In the limits $N\rightarrow\infty$ and $\Omega_{BS}\rightarrow \infty$, where $\nu = const$, the two cases are of the special interest:
the high temperature case, where $\bar\rho_{0}\,  e^{\mu\,Q\beta} < 1$ (warm BS), and
the low temperature case, where $\bar\rho_{0}\, e^{\mu\,Q\beta} \sim 1$ (cold BS).

In the early stage of the FBS (warm BS), the function $\bar n_{f}$  in the l.h.s. of (\ref{e15}) is regular on $f$, and the sum $\sum_{f} \bar n_{f}$ is replaced by the integral
where the spectrum of $f$ is continuous at $\Omega_{BS}\rightarrow\infty$:
$$\frac{1}{\nu} = \frac{1}{\Omega_{BS}}\,\sum_{f}\bar n_{f}\rightarrow \frac{1}{(2\,\pi)^{3}}\int\bar n(f)\,d^{3} f. $$ Using (\ref{e15}) one can find the formal equality:
\begin{equation}
 \label{e16}
 \int_{0}^{\infty}\frac{x^{2}\,dx}{\bar\rho_{0}^{-1}\,e^{-\mu\,Q\beta}\,e^{x^{2}} -1} = \frac{2\,\pi^{2}}{\nu}\,{\left (\frac{\beta}{2\,\mu_{s}}\right )}^{3/2}. 
\end{equation}
In the case of warm BS, the integral in the l.h.s. of (\ref{e16}) increases when the temperature increases as well.
  Thus, at given $\nu$ and $T$, the (\ref{e16}) has the relevant solution regarding the factor $\bar\rho_{0}\,  e^{\mu\,Q\beta}$ when
  \begin{equation}
 \label{e17}
 \mu_{s} > \frac{2\pi\beta}{(\nu B)^{2/3}}
\end{equation}
that comes from 
 $$\frac{2\,\pi^{2}}{\nu}\,{\left (\frac{\beta}{2\,\mu_{s}}\right )}^{3/2} <  \int_{0}^{\infty}\frac{x^{2}\,dx}{e^{x^{2}} -1},$$
where $B = 2,612...$ in (\ref{e17}) is the Riemann's zeta-function, $\zeta (3/2)$.  The scalar particles become massless when $T\rightarrow\infty$.
The case with warm BS is realised  when $T$ exceeds the critical temperature 
$$T_{c} = \frac{2\,\pi}{\mu_{s}} \left (\nu\,B \right )^{-2/3},  \,\,\,\,\mu_{s} \neq 0$$
in the case of the limits $N\rightarrow\infty$, $\Omega_{BS}\rightarrow\infty$ at
$$\nu\sim {\left (\frac{M^{\star}}{M_{Pl}^{2}}\right )}^{3}\frac{1}{N} = const.$$  
Here, $M^{\star}$ is the BS mass, $M^{\star}\sim M_{Pl}^{2}/\mu_s$, $\Omega_{BS}\sim (M^{\star}/M_{Pl}^{2})^3$ in the model of moderately relativistic BS with the momenta of the scalars $p\sim \mu_s$ [1],  $M_{Pl}\simeq 1.2\cdot 10^{19}$ GeV is the Planck mass. In the case of high enough $T$ the mass of the BS increases sharply ( as $\mu_s\rightarrow 0$) that leads to the possibility the BS may explode when $\nu\rightarrow\infty$ ($N\rightarrow 0$).

In the case of cold BS, when $\bar\rho_0\rightarrow 1$ and $T < T_c$, the small values of the  SDM momenta, $\vert p\vert \leq\delta$, are most important  for BEC. Here, 
\begin{equation}
 \label{e25}
\frac{1}{\Omega_{BS}}\sum_{\vert p\vert \leq\delta} \bar n_{p} = \frac{1}{\nu} - \frac{1}{\Omega_{BS}}\sum_{\vert p\vert \geq\delta} \bar n_{p} \simeq \frac{1}{\nu} - \frac{1}{(2\pi)^{3}} \int_{\vert p\vert \geq\delta} \frac{d^{3} p}{e^{E(p)\beta} -1}. 
\end{equation}
The result (\ref{e25}) is well-known, $\sim \nu^{-1}[ 1- (T/T_c)^{3/2}]$, where the only part of the total number of the particles proportional to $\sim (T/T_c)^{3/2}$ 
is distributed inside the BS with all the  momenta spectrum. The rest part, $\sim [ 1- T/T_c)^{3/2}]$, is the scalar condensate. 
The SDM particles with the mass $\mu_s$ (\ref{e17}) are taken to be sufficiently cold so that the particles may coalesce in the BEC state, the boson star.

Once the BS is formed, the particles are distributed in the volume $\Omega_{BS}$ at finite $T$ with some $p$ - momenta that affect the interactions between the particles and the decays. If we consider an arbitrary regular function $I(p)$ of the particle momentum in $\Re_{3}$ space, the average dynamical function $U$ related to the BS energy (mass) density $n^{\star}$ with $N$ scalar particles is 
$ \langle U\rangle /n^\star = N^{-1} \sum_{p} I (p) \bar n(p)$. On the other hand, $ \langle U\rangle /n^\star = \int I (p)\,\omega (p) d^3 p$, where for the cold BS 
($\bar\rho_0 e^{\mu Q\beta}\sim 1$) the $p-$ distribution function $\omega (p)$ of the particles contains the singular part,
\begin{equation}
 \label{e277}
\omega(p) = \left [ 1- {\left (\frac{\beta_c}{\beta}\right )}^{3/2}\right ]\delta (p) + \frac{1}{(2\pi)^3}\frac{\Omega_{BS}}{N} \frac{1} { e^{E(p)\beta} -1},\,\,\, \beta_c < \beta
\end{equation}
independently of an arbitrary function $I$.
In the vicinity of the critical point ($\beta_c \sim \beta$), the $\omega(p)$ is well-defined continuous function regarding the momentum $p$
\begin{equation}
 \label{e2777}
\omega(p) = \frac{1}{(2\pi)^3}\frac{\Omega_{BS}}{N} \frac{1} { \bar\rho_0^{-1}\,e^{F(p)\beta} -1}. 
\end{equation}
The thermo-statistical operator $F_{ths} (\beta; p, p^\prime)$ in $\Re_{3}$ is $F_{ths} (\beta; p, p^\prime) = \omega (p)\delta (p - p^\prime)$, while in $\Re^3$  the $F_{ths}$ is 
$F_{ths} (\beta;\bar r) = \int \omega (p) e^{i p\bar r} d^{3} p$, $\bar r = r - r^{\prime}$. Thus, at hight $T$ one has 
$$F_{ths}(\beta;\bar r) = \frac{\Omega_{BS}}{N} \int \frac{e^{ip\bar r}}{ \bar\rho_0^{-1}\,e^{F(p)\beta} -1} d^3 p,\,\,\, \beta_c < \beta, $$
and 
$$F_{ths}(\beta;\bar r) = 1 - {\left (\frac{\beta_c}{\beta}\right )}^{3/2} + \frac{\Omega_{BS}}{N}\int \frac{e^{ip\bar r}}{ e^{E(p)\beta} -1} d^3 p, \,\,\,\, \beta_c > \beta. $$
One can conclude that
$ F_{ths}(\beta;\bar r)\rightarrow 0$ as $\vert \bar r\vert \rightarrow \infty$ at $\beta_c < \beta,$ and 
$ F_{ths}(\beta;\bar r)\rightarrow  1 - {\left ({\beta_c}/{\beta}\right )}^{3/2}$ as $\vert \bar r\vert \rightarrow \infty$ at $\beta_c > \beta. $ Thus, we have a standard "jump" when going across $T_c$. This "jump" is nothing other but the BEC or the phase transition of the second kind. From ({\ref{e277}) we see that some part of the particles were "freezing" at $p = 0$. The number of these particles is defined by the relative quantity $\eta \sim 1 - {\left ({\beta_c}/{\beta}\right )}^{3/2}$ and this number is rather small at $\beta_c \sim \beta$. However, $\eta$ is increasing when the temperature is going down. At very low $T$ all the particles "freeze" and $\omega (p) = \delta (p)$. At $ 0 < T \leq T_c$ the BE condensate of almost "freezing" particles with the density $\hat\rho _s = \nu^{-1}\eta$  and the standard component of the particle density $\hat\rho_n = \nu^{-1}{\left ({\beta_c}/{\beta}\right )}^{3/2}$ are co-exist simultaneously and $\hat\rho_s + \hat\rho_n = N/\Omega_{BS}$. Note that the BEC is "visible" in $\Re_3$ momentum space. In the $\Re^3$ (coordinate) space, both $\hat\rho_s$ and $\hat\rho_n$ are mixing in such a case that one has the spatial homogeneity in case of the thermal and statistical equilibrium.

The fluctuations of particles inside the BS depend on the temperature.
We suppose that the particles are in the local volume $V$ of the BS bath with $\Omega_{BS}$. In the interior of the volume $V$ the scalar particles distribute almost uniformly at the density $\rho$.
The number of particles $n_{V}$ in $V$ is $\sum_{1\leq j\leq N} \hat n_{V} (q_{j})$, where $\hat n_{V} (q) = 1$ if $ q\in V$, and $\hat n_{V}( q) = 0$ otherwise.  The volume $V$ is defined by the geometry of the BS. 
The event-by-event fluctuation of the particle density $\langle {(n_{V} - \langle n_{V}\rangle )}^{2}\rangle $ with temperature is 
\begin{equation}
 \label{e27}
\frac{\langle {(n_{V} - \langle n_{V}\rangle )}^{2}\rangle}{ \langle n_{V}\rangle} -1 = \frac{\sqrt{2}\,\nu}{\pi^{2}} \left (\mu_{s}\,T\right )^{3/2} \int_{0}^{\infty} \frac{ x^{2}dx}{(\bar\rho_{0}^{-1}e^{-\mu Q \beta} e^{x^{2}} -1)^{2}} , 
\end{equation}
where $\langle n_{V} \rangle = V/\Omega_{BS}$. 
In the stage of the phase transition  when $\bar\rho_{0}\sim O(1)$ at $T\sim T_c$, the distribution (\ref{e27}) has no free parameters and may have the infinitely increasing behaviour as
$$ \sim \frac{4}{\sqrt{\pi}\,B} \int_{0}^{\infty} \frac{ x^{2}\,dx}{(e^{x^{2}} -1)^{2}}. $$
There is a stability of the BS at the stationary energy density point $\bar\rho_0$ where the BS as a quantum state is already formed. Having in mind that the number density of the BS is (see (\ref{e15}))
$$ n^\star\sim {\left (\frac{M^2_{Pl}}{M^\star}\right )}^3 \sum_{f} \frac{1}{\bar{\rho_0}^{-1} e^{F(f)\beta} -1}$$
and taking into account that $n^\star << 10^{19}\, pc^{-3}$ [4], one can find the condition for the stability of the BS at high enough $T$ for the maximal mass of the BS $\geq 10^{58}$ GeV: $\bar\rho_0 << O (10^{-16})$. The density $n^\star$ goes down rapidly at the phase transition, $n^\star\sim (2\pi)^3\,N (M_{Pl}^{2}/M^\star)^6\,\sum_{p} \omega (p)$ with $\omega(p)$ being in (\ref{e2777}).


\section{The DM observable rate with new scales}
We suppose the DM is embedded in particle physics, and thus, the new scales are necessary.
At very high energies the SM and the almost massless DM scalar particles can interact each other. This interaction may be very weak, suppressed by the powers of the "heavy" scale $M$ associated with the mass of "heavy" messenger (see., e.g., [10] for the case of  "unparticles").
The origin of the latter is unknown and we only flow below $M$ to analyse the structure of  interactions in the operator form
$$ \sim \sum_{n,l;k}^{n+l = 4 +k} \frac{O^{(n)}_{SM}\,O^{(l)}_{DM}}{M^{k}}, $$
where $O_{SM}^{(n)}$ and $O_{DM}^{(l)}$ denote the operators of the SM and the DM of the canonical dimensions $n$ and $l$, respectively; $k$ is the c-number associated with the scaling dimension of $M$.
We suppose that the leading interaction of the dynamical scalar field $\tilde S$ with the SM is through the Higgs portal.
Below $M$ the main contribution to the interaction potential is assumed to be in the ultra-violet form 
\begin{equation}
\label{e28}
\sim M^{2-\tilde d}\, {\vert H\vert}^2\, \tilde S,
\end{equation}
where $\tilde d$ is the scaling dimension of the operator $\tilde S$.
If $M$ is large enough, the field $\tilde S(x)$ does not couple strongly to the Higgs doublet ${\vert H\vert}^2$. 
The BS may be formed under the condition of the balance between the interaction (\ref{e28}) accompanied by the self-interaction of $\tilde S$ with the repulsive self-coupling from the one side, and the gravitational forces from another one (it will be considered in the Sec. 5). To that end, we lower $M$ to those level  at which the strength of the coupling between ${\vert H\vert}^2$ and  $\tilde S$ can be raised at some scale $\Lambda < M$. Below $\Lambda$ the coupling (\ref{e28}) flows to 
$\sim \xi (m^{2}/\Lambda) {\vert H\vert}^2 S$, where $S$ is the scalar operator of the scaling dimension $d$ in the non-trivial SI sector of an effective field theory, $m$ is associated with the Higgs boson mass.
 The coupling $\xi$ of the scaling dimension $(1-d)$ is strong enough as the powers of $\Lambda$ and $M$. When the SM Higgs gets the vacuum expectation value (VEV) $v$, the operator $S$ breaks the scale invariance with $\xi$ being in the form 
\begin{equation}
\label{e29}
\xi \sim \Lambda^{\tilde d - d +1}\,M^{2 - \tilde d}\, \frac{1}{m^{2}}.
\end{equation}
The SDM  sector and the SM scalar sector are separated by the scale 
$$ \tilde\Lambda = \left (\xi\,\frac{m^{2}}{\Lambda}\,v^2\right )^{\frac{1}{4-d}} < \Lambda $$
below which
the SDM sector becomes the SM sector. Thus, one can define the energy barrier  $E_{\star} \geq \tilde\Lambda$ at which the BS can exist. The interaction between the SM and the SDM is very weak when $\tilde d\rightarrow 1$,  $\ d\rightarrow 1$ and $M\sim \Lambda\sim O(v)$. Any (observable) quantity $\varepsilon_{\star}$ related to the BS is restricted in the sense to be observed at the level 
$$ \varepsilon_{\star} = {\left [ \xi\,\frac{1}{M^{n-2}}\,\frac{m^{2}}{\Lambda}\, E_{\star}^{d+n-4}\right ]}^{2}. $$
Taking into account $E_{\star} \geq \tilde\Lambda$ the observable $\varepsilon_{\star} $ is bounded by the ratio between $E_{\star}$ and $M$ scaled with $n$ only 
$$\varepsilon_{\star}  \leq {\left (\frac{E_{\star}}{M}\right )}^{2n}  {\left (\frac{M}{v}\right )}^{4}. $$
For $n = 4$ we find $\varepsilon_{\star}  \leq 10^{-61}$ if $E_{\star}$ is of the order of the conformal symmetry breaking scale $E_{\star}\sim f \geq 3$ TeV [11,12] and $M\sim M_{Pl}$. So, the effect of an observation of the deviation from the SM within the coupling (\ref {e28}) is rather small at the Planck scale. However, if we suppose to have the effect on the level of $\sim 1 \%$, the scale $M$ should be as low as $\sim 100$ TeV that would be an effective test for the FCC-hh. 

Assuming the interaction between the SM and the dark sector mediated by the operators of the dimension $\tilde d +2$ with $\tilde d \geq 2$, the production cross section scales with the energy $E$ as $\sigma \sim \omega^2 {\left (E/M\right )}^{2 (\tilde d - 3)}, $
where $\omega$ is the dimensionless constant. 
Looking through the interaction (\ref{e28}), the production rate of the SDM at the $O$(GeV) scale at the colliders is $N_{coll} = L_{int}\,\sigma\,\varepsilon_{\star}$, where $L_{int}$ is the integrated luminosity.
One can compare the results for new physics (NP) with the SDM discovery if it is possible to reach of a $\sqrt s_{FCC} = 100 $ TeV proton-proton collider FCC-hh at $L_{int}$ = 30 ab$^{-1}$ with the one of the HL-LHC at a $\sqrt s_{LHC} = 14 $ TeV with $L_{int}$ = 3 ab$^{-1}$ (the review with the NP discovery at the FCC collider can be found in [13]). The relative production rate is 
$$ r_{FCC/LHC} = 10\times {\left (\frac{\sqrt {s}_{FCC}}{\sqrt {s}_{LHC}}\right )}^{2 (\tilde d - 3)} \times {\left (\frac{\tilde\Lambda_{FCC}}{\tilde\Lambda_{LHC}}\right )}^{2n}. $$
Since $\tilde\Lambda \leq E_{\star} \leq \sqrt s$, the effect of the advantage of the FCC-hh would be as $ r_{FCC/LHC} \simeq 10\times 50^{\tilde d + n - 3}$ that is  $ r_{FCC/LHC} \simeq 500$ in the minimal case of the interaction (\ref{e28}) with the scaling dimension $\tilde d = 2$ of the operator $\tilde S$  and the SM dimension $n = 2$.

\section {The Model. BS potential} 
In the early Universe with the  temperature $T_{univ}$ the fields of particles known from the SM and the DM sector are almost free and if they interact each other, that only feebly according to the SI symmetry principle at the scale $\Lambda \leq T_{univ}$.  An intriguing story in the evolving Universe is related  with scalar fields, both in the SDM and the DM sectors. 
The SI is violated by the presence of $M_{Pl}$, $\Lambda$ and the VEV's of the Higgs and the SDM. These scales may be dynamical if they are replaced by the scalar fields. In particular, 
one can introduce the different particle physics scalar fields $\Phi (x)$ in the Einstein-Hilbert term for gravity in the corresponding action by means of the following approximation: $R/G \sim const \left (\sum_{\Phi} a_{\Phi} \Phi^{n}\right ) R + ...,$
where $R$ is the Ricci scalar for the background metric $g_{\mu\nu}$, $G = (8\pi M_{Pl}^2)^{-1}$, $a_{\Phi}$ is a new parameter, $n$ stands for the integer number. The model action for the Higgs doublet ${\vert H (x)\vert}^2$ and the singlet SDM field $S(x)$  reads
\begin{equation}
\label{e3722}
 A = \int d^{4} x \sqrt {-g} \left [\left (a_h {\vert H\vert }^2 + \frac{1}{2} a_{s} S^2 \right ) R - \frac{1}{2} g^{\mu\nu}\partial_{\mu}S\partial_{\nu}S - V(S,  {\vert H\vert}^{2})\right ], 
 \end{equation}
where $a_h$ and $a_s$ are real positive constants,  $g$ is the determinant of $g_{\mu\nu}$; $\Lambda \leq \Lambda_{inf} \sim O(10^{16} GeV)$ with $\Lambda_{inf}$ being the scale of the Universe inflation. For concreteness, we consider the potential $V$ in (\ref{e3722}) in the form 
\begin{equation}
\label{e317}
V  (S, {\vert H\vert }^2) =  \frac{1}{2}\lambda {\left ( {\vert H\vert }^2 - \alpha^2 S^2\right )}^2 + \frac{1}{4}\zeta\, S^4,
\end{equation}
where $\lambda,\alpha, \zeta$ are the constants. The action (\ref{e3722}) with (\ref{e317}) is scale-invariant under the global transformations $g_{\mu\nu}(x)\rightarrow g_{\mu\nu}(\kappa x)$, 
$\Phi (x)\rightarrow \kappa^d\Phi (\kappa x)$, where $\kappa$ is an arbitrary real parameter.
An additional $Z_2$ symmetry assumes the $S$-particle is stable  and is viable candidate for scalar DM.
Neglecting the gravitational part in (\ref{e3722}) in the non-flat direction ($\zeta\neq 0$) for $\alpha^ 2 > 0$ and $\lambda > 0$, the family of the ground states is defined through the equations
$$ 2 H^+H = v^2, \,\,\, v^2 = \frac{ 2\lambda\alpha^4 + \zeta}{\lambda\alpha^2} s_0^2,$$
where $s_0$ is an arbitrary real constant relevant to the field $S(x)$. These ground states  are modified when the gravitational interaction is included:
\begin{equation}
\label{e3177}
 v^2 = 2\,\alpha^2 s_0^2 + \frac{2}{\lambda} a_h R,\,\,\,\,\, R = \frac{\zeta}{a_s + 2\,\alpha^2 a_h} s_0^2. 
 \end{equation}
The hierarchy between the SDM scale and the gravitational scale in terms of the ratio $s_0^2/M_{Pl}^2$ depends on a set of the coupling constants $a_s, a_h, \lambda$ and $\alpha$ in the non-flat direction, and does not depend on $\lambda$ in the flat space-time ($\zeta = 0$)
\begin{equation}
\label{e3171}
\frac{s_0^2}{M_{Pl}^2} = \frac{1}{a_s + 2a_h\left ( \alpha^2 +\frac{a_h}{\lambda}\frac{\zeta}{a_s + 2\alpha^2 a_h}\right )}.
\end{equation}
From the physical point of view with $(s_0/M_{Pl})^2$ being of the order $\sim O(10^{-32})$, 
the coupling constants $a_s$ and $a_h$ in (\ref{e3171}) have to take values satisfying $a_s, a_h >> 1$. On the other hand, the ratio (\ref{e3171}) may be presented in the form 
\begin{equation}
\label{e3173}
 \frac{s_0^2}{M_{Pl}^2} = \frac{1}{a_s }\left (1 - a_h\frac{v^2}{M_{Pl}^2}\right ) 
 \end{equation}
from which $a_h \leq O(10^{33})$ that indicates the strong coupling between the Higgs boson and the gravity. For the scale of the SDM $s_0\sim f\sim O(10^3)$ GeV [12], the coupling $a_s$ is bounded by 
\begin{equation}
\label{e3172}
a_s \geq O(10^{32}\Delta_{\epsilon_\star}),
\end{equation}
where $\Delta_{\epsilon_\star} = 1 - \epsilon_\star$ and $\epsilon_\star = a_h \left (v^2/M_{Pl}^2\right ) \leq 1$. The result (\ref{e3172}) gives the SDM/Higgs - gravity seesaw coupling mechanism when the increasing of the coupling $\sim {\vert H\vert }^2 R$ causes the decreasing of the coupling  $\sim S^2 R$ and vice-versa. 
In the flat direction, from (\ref{e3171}) and (\ref{e3173}) one has $\alpha^2 \leq O(10^{-2})$. The cosmological constant $\Lambda_{cosm}$ is associated with the term $\sim (1/4) \zeta S^4$   in (\ref{e317}), from which one can find 
$$\Lambda_{cosm} \simeq \frac{1}{4} \zeta \frac{M_{Pl}^4}
{\left (a_s + 2 \,\alpha^2 a_h\right )^2} {\left ( 1 - \frac{4 a_h^2}{\lambda} \frac{\zeta}{a_s + 2\alpha^2 a_h} + ...\right )}^2 $$
and approximately
$$\frac{\Lambda_{cosm}}{M_{Pl}^4}\sim  \zeta O\left (10^{- 64} \right ),$$
where $\zeta << 1$ (see (\ref{e3177})) and $\lambda \leq O(1)$ as corresponding to the self-coupling of the Higgs boson field.


At  $T_{univ} < \Lambda$ the breaking of the SI may be regulated by discrete manner, where the BS may contain the massive SDM fields $\hat S$ considered here in the "tower" form with an arbitrary scaling dimension $d$,
$\hat S(x) = \sum_{k=1}^{N} c_k\, \varphi_{k}(x) $.
The coefficients $c_k$ depend on the dynamics of underlying conformal field theory (CFT). By assumption, this dynamics must be such that the model potential is minimised at $\langle \varphi_k\rangle = f_k$, where $f_k$ is the order parameter for scale symmetry breaking determined by the  dynamics of the hidden strong sector. In general, the $c_k$ are functions of the scaling dimension $d$.
 The coefficient $c_k$ for each $k =1,2,..., N$ acts as the conformal regulator controlling the "weight" of the conformal mass parameter $\Delta^{2}$ scaled as  $k\,\Delta^{2} /N \sim \mu_{k}^{2}$ in the "tower" $\hat S(x)$  of the light scalar states  $\varphi_k(x)$ with the mass $\mu_k$. 
We assume that $\mu_k$ is much smaller than other scales in the effective model.
If $\Delta^2$ is small but finite, the SDM can decay, and we shall discuss of such decays.
The model matches that conformal symmetry is recovering in the limit $\Delta^2\sim\mu_k^2 \rightarrow 0$.
It is not possible to make detailed predictions without knowledge of the coefficients $c_k$ in the operator $\hat S$. To clarify this item, we consider 
the propagation of the $\hat S(x)$ which is governed by the function [14,15]
$$ D(p^2, \mu_k; c_k;N) = \sum_{k=1}^{N} \frac{{\vert c_k\vert}^2}{p^2 -\mu_{k}^2 + i\epsilon}, $$
where ${\vert c_k\vert}^2$ is related with 
 $\varrho (t,\mu_k; c_k;N)$  as $\varrho(t, \mu_k; c_k;N) = \sum_{k=1}^{N} {\vert c_k\vert}^2 \,\delta (t - \mu_{k}^2).$
Going to the following form  of the propagator
\begin{equation}
\label{e35}
  D(p^2) = \int_{0}^{\infty} \frac{dt}{2\pi} \frac{\varrho(t)}{p^2 - t + i\epsilon} = \int d^4 x\, e^{ipx} \langle 0\vert T \hat S(x) \hat S^{+}(0)\vert 0 \rangle 
\end{equation}
in the limit $k\rightarrow\infty$, one can find $ {\vert c_k\vert}^2 = const\, \Delta^{2} {\left (\mu_k^2\right )}^{d-2}   $
for the SI requirement that $\varrho (t)$ has to be as $\sim t^{d-2}$, $ d \geq 2$.  
In the field theory, $\hat S (x)$ can also be constructed by convoluting the scalar field $\varphi (x,t)$ with a function $c(t)$ with a fixed scaling dimension $d$ to have the following form 
$$ \hat S(x)\rightarrow S_d(x) = \int_0^\infty c(t) \varphi (x,t) dt,$$
where $t$ is a continuous mass squared parameter, and $c^2(t)\sim t^{d-2}$. The generalised field $S_d(x)$ has the same phase space, the spectral density and the propagator as that the "tower" operator $\hat S$. 

The SI extension of the SM plus gravity including the SDM in the "free scale " form $\sim \sum_{k=1}^N\varphi_k^2 (x)$ is given by the action
\begin{equation}
\label{e371}
  A _{SI}= \int d^{4} x \sqrt {-g} \left [ \left (a_h {\vert H\vert }^2 + \frac{1}{2} a_{\varphi} \sum_{k=1}^{N} \varphi_{k}^2 \right ) R  - V(\varphi_k; {\vert H\vert}^{2})\right ], 
 \end{equation}
where 
\begin{equation}
\label{e373}
 V(\varphi_k; {\vert H\vert}^{2}) =\frac{1}{2} \lambda_1\left ({\vert H\vert}^{2} -\frac{1}{2}\alpha_1^2 \sum_{k=1}^N\varphi_k^2\right )^2 +\frac{1}{4}\zeta_1 \sum_{k=1}^N\varphi_k^4
 \end{equation}
with $a_\varphi$, $\lambda_1$, $\alpha_1$ and $\zeta_1$ being the real positive constants. The inclusion of gravity and the Higgs - SDM interactions leads to new additional  ground state
\begin{equation}
\label{e380}
v^2 = \left (\alpha_1^2 + \frac{2 a_h}{\lambda_1} \frac{\zeta_1}{a_\varphi + \alpha_1^2\, a_h}\right ) \sum_{k=1}^{N} f_{k}^2.
\end{equation}
At the scales $\leq\Lambda$  the SI is broken  
with the main interactions  $\sim \hat S {\vert H\vert }^2  R$ and $\sim {\vert H\vert }^2 \hat S$  which can allow to extract the coefficients $c_k$ in the scalar "tower" $\hat S(x)$. 
The action (\ref{e371}) is replaced with $A_{SI}\rightarrow A = A_{SI} + A_\Lambda$, where 
\begin{equation}
\label{e372}
  A _\Lambda = \int d^{4} x \sqrt {-g} \left [ \frac{\left (a_{\varphi h} R - \xi m^2\right ) }{\Lambda} {\vert H\vert }^2 \hat S  - 
  \frac{g^{\mu\nu}}{2\, \Lambda^{2(d-1)}}  \partial_\mu \hat S \partial_\nu \hat S - V_1(\varphi_k; {\vert H\vert}^{2})\right ].
\end{equation}
 Here,  $\partial_\mu \hat S(x) = \sum_{k = 1}^{N} c_k \partial_\mu\varphi_k (x)$, 
 $a_{\varphi h}$ is the parameter  of the scaling dimension $(1-d)$ and the potential $V_1$ is
 \begin{equation}
\label{e3772}
 V_1(\varphi_k; {\vert H\vert}^{2}) = - m^2 {\vert H\vert }^2 + \frac{\beta^2}{2} \zeta_1 \left ( \frac{\beta^2}{2}\sum_{k=1}^{N} f_{k}^4  -  \sum_{k=1}^{N} f_{k}^2\, \varphi_k^2 \right ), 
 \end{equation}
 where 
 $\beta^2\simeq 1+ O(1/N)$ with $N$ being the number of particles in the $\hat S(x)$ "tower".  In the limit of exact SI the SDM field $\hat S(x)$ may be derivatively self-coupled, $\sim c_4 (\partial_\mu \hat S\partial^\mu \hat S)^2/ \hat S^4$. The inverse power of $\hat S^4$ is needed the Lagrangian density would have the correct transformations under the scalings. However, we do not use this self-coupling when $\hat S$ is presented in terms of the scalar "towers" and the constant $c_4$ is unknown and depends on the details of the hidden conformal sector. 
Since the $\hat S$ "tower" is treated as a non-local operator, 
there is no direct interactions of $\hat S(x)$ with the gravity, however this interplay is expected to be through the Higgs boson (the first term in (\ref{e372}).
There is no interactions of two SDM "towers" $\hat S \hat S $ with the gravity, with the Higgs boson ${\vert H\vert }^2$ and the self-interaction of $\hat S(x)$. 
If $\hat S(x)$ would be the local operator, the product $\hat S_i(x)\hat S_j(x)$ is properly normalised another operator $\hat S_l(x)$ of the dimension $d_l$ equal to the sum of dimensions of individual operators $\hat S$ labeled by $i$ and $j$, $d_l = d_i + d_j$ (see also [16]).
One can use instead the corresponding interactions of the SDM in terms of the real scalar local  fields $ \sum_{k=1}^{N} \varphi_{k}^2$ with the appropriate couplings.
We assume that the product of the couplings $\sim\lambda_1\alpha_1^2$ in (\ref{e373}) is not small so that the field $\varphi_k$ is in the thermal equilibrium with the Higgs boson in the bath at high $T$. As the temperature drops down to the mass scale $\sim \mu_k$, the heavier $\varphi_k$ freeze out and decouples from the thermal bath in the BS. Then, $\varphi_k$ after freeze out may decay into the photons and the dark photons. The DP would later decay to leptons or neutrinos. As a consequence,  the density of  cosmic rays could be compared with the (relic) density of the cosmic rays  collected from cosmological data.

In field theory, the operator of the SDM has to be considered in the continuum limit instead of the "tower" approximation $\hat S(x) = \sum_{k=1}^{N} c_k\, \varphi_{k}(x) $. To that end, one needs to calculate $f_k$ first using the minimisation condition for $\varphi_k (x)$. The result is 
\begin{equation}
\label{e40}
f_k = \left ( \frac{v^2}{\Lambda}\right )
 \frac{\xi\,m^2 - a_{\varphi h}\,R}{2 \,a_\varphi\, R + \lambda_1\,\alpha_1^2 \left [v^2 - \alpha_1^2\left ( \sum_{n=1}^N f_{n}^2 \right ) \right ] + 2 \,\zeta_1\, \mu_{k}^{2}}\cdot c_k,
\end{equation}
where $\mu_k^2 = (\beta^2 -1) f_k^2$.
The BS stability $f_k > 0$ has to be valid even for $\zeta_1\rightarrow 0$.  
The  $f_k$ (\ref{e40}) decreases when $\zeta_1$ goes away from the flat direction for arbitrary constants $a_{\varphi h}$, $a_{\varphi}$, $\lambda_1$ and $\alpha_1$.
One can easily find that the VEV of the field $\varphi_k (x)$ is governed by the VEV of the Higgs plus gravity, and no infra-red catastrophe is emerged if $\mu_{k}^2 \rightarrow 0$. 
The stability condition of the BS is given in terms of the VEV of the Higgs boson and $f^2$ plus gravity keeping  the VEV of the SDM operator $\hat S(x)$ in the continuum limit to be finite 
\begin{equation}
\label{e41}
\langle \hat S\rangle = \left \langle \sum_{k = 1}^\infty c_k \varphi_k\right \rangle \sim  \sum_{k = 1}^\infty c_k f_k \rightarrow
\langle S \rangle = \Lambda \left (\frac {f}{v}\right )^2\frac {m_{gap}^2}{\Delta_{Rm}} \frac {1} {2-d}, \,\,\,  1 < d < 2,
\end{equation}
where $f_k$ is given in (\ref{e40}).
The VEV $f^2$ is finite in the non-flat space-time ($\zeta_1 \neq 0)$ and non-zero $\Delta_{Rm} = \xi\,m^2 - a_{\varphi h}\,R$, 
\begin{equation}
\label{e411}
 f^2 = \left (\frac{v^2}{\Lambda}\right )^2 \frac{\Delta_{Rm}^2}{\left (m_{gap}^2\right )^{3-d}}\left (2\zeta_1\right )^{1- d} \Gamma(d-1)\Gamma (3-d),\,\,\,\, 1 < d < 3.
 \end{equation}
The strength of $f^2$ depends on the scaling dimension $d$ in the power of the gap mass quantity $m_{gap}^2 = 2\,a_\varphi\,R + \lambda_1\,\alpha_1^2 v^2\,\left [ 1 - {\left ( \alpha_1 f/ v \right )}^2 \right ]$.
In case the gravity is neglected,
the VEV (\ref{e41}) is finite, however, it flows to the scalar condensate when $\xi\rightarrow 0$ as $M\rightarrow\infty$ at $\tilde d >2$ (see (\ref{e29})). 
The new ground state includes (\ref{e380}) and has the form 
\begin{equation}
\label{e402}
 v^2 = \frac{2\, m^2}{\lambda_1} \left ( 1 - \xi\frac{\langle S\rangle}{\Lambda}\right ) +  \frac{\alpha_1^2}{\lambda_1} f^2 + \left (2 a_h + \frac{a_{\varphi h}}{\lambda_1} \frac{\langle S\rangle}{\Lambda}\right ) R.
\end{equation}
The value (\ref{e402}) determines the mass scales of particles and strongly influences the phenomenology at the scale below $\Lambda$ including the stability of the BS.
The VEV (\ref{e402}) shows how the model with the Higgs - SDM interactions where the gravity is also included can modify the SM relation. The Higgs mass is shifted away from the SM value $\sim m$.
The properties of the Higgs is changing already at the tree level making the BS as that the bound state where the Higgs and SDM are a mixed sector interacting with gravity. 
Actually, the result  for $v^2$ with (\ref{e402})  flows to the SM Higgs value if the couplings between ${\vert H\vert }^2$ and SDM are neglected. 
The (\ref{e402}) also shows how the approximate CFT admits the second ground state, where the history in the evolving Universe is conventional. The main explanation of this mechanism is a SDM in the few dozens GeV range which mixes with the Higgs and can be detected either at high energies colliders or in the cosmological experiments through the neutrino detection (Sec. 6).

At the cosmological scales $\sim O(M_{Pl})$ the strength of interactions between  the matter fields and the gravity is defined by the average values of the corresponding fields. In this case, one can identify $M_{Pl}^2$ as 
$$ M_{Pl}^2 = a_\varphi f^2 +\left ( a_h + a_{\varphi h} \frac{\langle S\rangle}{\Lambda} \right ) v^2. $$
The cosmological constant $\Lambda_{cosm}$ is associated with the term  (see ({\ref{e373}) and (\ref{e3772}))
$$\sim (1/4) \zeta_1 \sum_{k=1}^\infty \left (\varphi_k^2 - \beta^2 f_k^2\right )^2$$
 from which one can get  
 $ \Lambda_{cosm} \sim \zeta_1 \left (\beta^2 -1\right )^2 M_{Pl}^4/a_\varphi^2 \sim \zeta_1 \left (\beta^2 -1\right )^2 O(10^{-6})\,eV^4$ if the gravity is neglected.  Here, the constant $a_\varphi \sim O(10^{32})$.
 The two parameters  $\beta\sim O(1)$ and $\zeta_1 << 1$ are responsible for the very small ratio $\Lambda_{cosm}/M_{Pl}^4$.
In order to estimate the lifetime of the SDM particle at the cosmological scales,
we assume the approximate $Z_2$ symmetry in the model where the couplings $A_R$ and $A_\xi$ between $\hat S$ and the Higgs boson in $\sim (A_R + A_\xi) \hat S {\vert H\vert }^2$ are small enough. 
Here, 
\begin{equation}
\label{e43}
 A_R = \frac{a_{\varphi h}}{\Lambda}\,R,
 \end{equation}
while $A_\xi  = \xi  \Lambda^{-1} m^{2} << m_{h}$ for $\xi\sim O(1)$. 
The scalar curvature $R$ in (\ref{e43}) is $d$-dependent and defined from (\ref{e411}).
 The drivers of the formation of the BS when the latter could arise are the following:\\
 - the self-interaction of the scalar fields with the couplings $\lambda_1\alpha_1^4$;\\
 - in asymptotically flat space-time the gravitational equilibria of the scalar fields with the mass is characterised by 
 $R = \xi m^2/a_{\varphi h}$ that means $A_R = A_{\xi} $;\\
 - the course of gravitational condensation of scalar DM in the early Universe, $\langle S(x)\rangle\rightarrow \infty$.\\
More arguments  in favour of the case  $\zeta_1 = 0$ can be found in [17]. The lifetime $\tau_s$ of the $S$- particle is rescaled with that of the Higgs boson $\tau_h$ by mixing parameter $\kappa$ as $\tau_s\sim \kappa^{-2}\tau_h\sim 10^{12}$ seconds, where $\kappa = (1/2)\xi\,v/\Lambda$ and $\tau_h$ is the lifetime of the Higgs boson (see, e.g., [18], $\tau_h = 1.6\times 10^{-22}$ seconds).
The lifetime $\tau_s$ can already be compared with the age of the Universe $\tau_{univ} \sim 10^{17}$ seconds. 
The including of the gravity (the non zero term (\ref{e43})) leads to the fact the lifetime of the scalar dark particle becomes shorter and consequently, the dark photons and the direct photons will have less chance to be escaped. If the $S$-particle is stable ($\tau_s \geq \tau_{univ}$), the constant $\xi$ in the main interaction $\sim \vert H(x)\vert ^2 \hat S(x)$ should be small, $\xi \leq 3\cdot 10^{-3}$.
The long-lived SDM could coalesce into scalar DM halos which constitute to the formation of the Galaxy.

\section{Production of the SDM and its decay }

The interaction between the SDM with the spin 1/2 DM $\chi_i$ is of the following form 
\begin{equation}
\label{e351}
 \frac{c}{\Lambda^d}\,\bar\chi_1\,\gamma^{\mu} (1 - \gamma_5)\chi_2\,\partial_\mu \hat S + h.c.
\end{equation} 
Here, $c$ is a dimensionless constant, $\partial_\mu \hat S(x) = \sum_{k=1}^{N} c_k\,\partial_\mu \varphi_ k (x)\simeq B_\mu (x)$ is the DP field $\bar\gamma$ [8]. The coupling (\ref{e351}) may assume the production of $\varphi_k$ in the transition $\chi_2\rightarrow\chi_1 + \varphi_k$, where the mass of $\chi_2$ is bigger than that of $\chi_1$. 
Because the BS is not absolutely stable depending on the lifetime of the SDM inside, the dark scalar particles
may decay to the photons (the primary or direct photons) or/and to dark photons $\bar\gamma$ where the latter  may subsequently decay into the lepton pairs $\bar l l$
with the couplings $\sim c_A\,\Lambda^{1-d}\,\bar l\,\gamma^{\mu}\gamma^{5}\,l \,B_{\mu}$ with $c_A$ being the appropriate constant.
The range of the kinetic mixing strength $\varepsilon$  between the $\bar\gamma$ and the ordinary photon  is from $\sim 10^{-12}$ in the cosmological study of the DM signals [19] raising to level of $\sim 10^{-4}$ in the experiments at high-energy intensity frontier [20] up to $\sim 10^{-2}$ in the search for DP at the accelerator-based experiments [21]. 
The actual detectability depends on the scale $\Lambda$, the constant $c_A$ and the mixing strength $\varepsilon$. One can look at the charged leptons tracks because the daughter particle has a fixed energy, and/or (missing) energy measurement.

The propagator $ D^{\mu\nu}(p^2) =  \int d^4 x\, e^{ipx} \langle 0\vert T B^{\mu}(x) B^{\nu}(0)\vert 0 \rangle$ of the field $B_\mu (x)$ has the similar form as those in (\ref{e35})
   $$ D^{\mu\nu}(p^2) = \left ( g^{\mu\nu} - \frac{p^\mu\,p^\nu}{p^2}\right ) \int_{0}^{\infty} \frac{dt}{2\pi} \frac{\varrho(t)}{p^2 - t + i\epsilon} $$
with $\varrho (t) \sim t^{d-2}$. The DP field $B_{\mu}(x) = \sum_{k=1}^{N} \tilde c_k\,\bar\gamma_{\mu,k}$ is defined through the matrix element $\langle 0\vert B^{\mu}(0) \vert \bar\gamma_k\rangle = \epsilon^\mu \tilde c_k$, where $\epsilon^{\mu}$ is the polarisation of the DP $\bar\gamma_k$. At finite $k$ the propagator is 
$$ D^{\mu\nu}(p^2, \tilde\mu_k; \tilde c_k;N) = \left ( g^{\mu\nu} - \frac{p^\mu\,p^\nu}{p^2}\right )\sum_{k=1}^{N} \frac{{\vert \tilde c_k\vert}^2}{p^2 -\tilde\mu_{k}^2 + i\epsilon}, $$
where the DP mass $\tilde\mu_{k}^{2} \sim \Delta^{2}\cdot k$ and ${\vert \tilde c_k\vert}^2\sim \Delta^{2} {\left (\tilde\mu_{k}^{2}\right )}^{d-2}$. 

In case of the visible decay of the DP, the lifetime $\tau_{\bar\gamma}$ of the latter if measured through the decay length (the mean displacement of the event vertex), will be scaled with $\Lambda$ and the DP mass $\tilde\mu$, and increased with $\varepsilon^2$:
\begin{equation}
 \label{e341}
\tau_{\bar\gamma} \simeq {\left (\frac{\Lambda}{\tilde\mu}\right )}^{2(d-1)}\,\frac{3}{c_{A}^{2}\,\alpha\,\varepsilon^{2}\,\tilde\mu\,\Re_{eff}}. 
\end{equation}
Here, $\alpha$ is the electromagnetic coupling constant, $\Re_{eff}$ is the effective number of the possible decay channel, either $\bar\gamma\rightarrow e^+e^-$, or $\bar\gamma\rightarrow \mu^+\mu^-$. 
The constraints on $\varepsilon$ depending on the DP mass were obtained by BABAR [22] and NA64 [23] experiments, leading to $\varepsilon \leq 10^{-3}$ for the dark photon mass $\leq 8 $ GeV.
If $\tilde\mu$  does not exceed two masses of the muon, $ 2 m_{\mu}$, $\Re_{eff} = 2$. For  $\tilde\mu\geq 2 m_{\mu}$ one has 
$ \Re_{eff} = \left ( 1+ A_{\mu^+\mu^-} \right) \left [1 + R _{(had/\mu^+\mu^-)}\right ]$, where 
$A_{\mu^+\mu^-} = \sqrt {1 - \left (2 m_{\mu}/\tilde\mu\right)^2} \left [1 +  2\left ( m_{\mu}/\tilde\mu\right)^2\right ]$ and $R _{(had/\mu^+\mu^-)}$ being the ratio of the $e^+e^-$ annihilation into hadrons and for the $\mu^+\mu^-$-pair production
$$ R _{(had/\mu^+\mu^-)} = \frac{\sigma (e^+e^-\rightarrow hadrons)}{\sigma (e^+e^-\rightarrow \mu^+\mu^-)} = \frac{6\pi}{\alpha}\,g_{\varphi\gamma\gamma}\,f$$
with the coupling between the scalar field $\varphi$ and two photons $g_{\varphi\gamma\gamma} \leq 10^{-6}\, GeV^{-1}$ defined by the conformal anomaly term in the interaction between the conformal sector and the SM [8]. 
For concreteness, if $d\rightarrow 1$ and $c_A \sim O(1)$, one has  $\tau_{\bar\gamma} \geq 10^{-26}$ seconds. On the other hand, if one admit that $d\rightarrow 2$, then 
the DP becomes almost long-lived boson with the life-time  $\tau_{\bar\gamma} \geq 10^{10}$ seconds at the cosmological scale $\Lambda \sim O(M_{Pl})$. 
The estimation of $\tau_{\bar\gamma} $ within (\ref{e341}) has no the dependence of available decay channel, because of the small contribution in $\mu^+\mu^-$ channel, $R _{(had/\mu^+\mu^-)} \sim O(10^{-2})$.
The BABAR collaboration [24] searched for bound states composed of the dark fermions ("darkonium") where the latter decays to 3 dark photons, and these DP's decay to electrons, muons, or pions. 
No significant signal is observed for the DP lifetimes $\tau_{\bar\gamma}\sim 10^{-12}$ sec, $\tau_{\bar\gamma}\sim 10^{-11}$ sec, and $\tau_{\bar\gamma}\sim 10^{-10}$ sec for the DP mass  below 0.2 GeV and the limits on the cross section for each value of $\tau_{\bar\gamma}$ were reported. 

At the early stages of its evolving, the Universe may be transparent for free degrees of freedom of the strong interacting matter. Let us assume that the FBS is due to SDM particles and the Higgs bosons produced in the fusion of gluons. The interaction between the SDM and the gauge boson field defined by the gluon field strength tensor $G_{\mu\nu}^a$ is 
$\sim (\alpha_s/8\pi) C_g (S/\langle S\rangle) (G_{\mu\nu}^a)^2$, where $C_g = 11 - (2/3)n_{light}$ with $n_{light}$ being the number of quarks lighter than the $S$-particles [25]. The significance of the SDM contribution to the FBS to that of the Higgs bosons is 
\begin{equation}
 \label{e3411}
\frac{Q_S}{Q_{Higgs}} = \left (C_g \frac{v}{\xi\,\langle S\rangle}\right )^2 
\simeq \left \{\frac {C_g\, m^2 v\,(2-d)}{\Lambda f^2 \lambda_1\alpha_1^2 \left [1 - \left (\alpha_1 f/v \right )^2\right ]} \right \}^2,
\end{equation}
where $ 1 < d < 2$ and  the gravitation is neglected. At the cosmological scales one has the following lower bound on the significance (\ref{e3411}): $Q_S/Q_{Higgs} > O(10^{-34})$ for $\lambda_1 \sim O(1)$, $ d\rightarrow 2$, $\alpha_1^2 \leq O(10^{-2})$ and $f = 3$ TeV.

In the end of this section we consider the decay of the SDM inside a BS into SM neutrinos ($\nu$) where the latter were resulted with the Super-Kamiokande (SK) detector. Using the SK results, the maximal distance $r_{\nu}$ away from the $\nu$ source inside the BS to the detector on the Earth can be estimated (see also [26]). The $r_\nu$ constraint depends on the SDM mass $\mu_s$, its lifetime $\tau_s$, the data taking time and the effective area of the detector. The SK reduced limit on the number of events within the data taking for 1679.6 days is 
$$N_{SK}^\nu = 10^{31} \left (\frac{10^{18} sec}{\tau_s}\right ) \left (\frac{10\, GeV}{\mu_s}\right )^{5/4} \left (\frac{D_{SK}}{4}\right )^2\frac{1}{r_{\nu}^2} \leq 1,$$ 
where $D_{SK} = 33.8$ m is the SK linear size of the effective detection area. The constraint on the distance is  $r_\nu\leq 1$ kpc for $\tau_s\sim 10^{12}$ sec (see Sec. 5) and the lower bound on SDM mass $\mu_s\geq 60$ GeV. The $\nu$ emitting source in the BS is located in our Galaxy and is about 8.5 times closer to the Earth compared to that of the 
Galactic Center distance of the solar system ($r_{\odot} = 8.5$ kpc).  


\section{Conclusion}

In conclusion, the BS would remain a good and an instructive laboratory for testing the ideas of interactions between the SM and  the dark scalar sector plus gravity.  This is owned to the discovery the scalar boson stars behave as the large systems with the mass $\sim O(M_{Pl}^{2}/\mu_s)$ for the dark scalar mass $\mu_s$.
One of the purposes of this paper is to clarify the scalar dark matter using the language familiar to the quantum field theory.
The SDM properties emerge through the concrete interactions.
In the paper, we have investigated the couplings of the Higgs to a dark scalar operator $\S$ of the scaling dimension $d$. 
The SDM can play the role of a fundamental field with the VEV $\langle S \rangle$ in the late stage of the evolution of the Universe during which the SDM field is going down to the stable potential valley. 
We find that at lower $T$ and at the scale $< \Lambda$ the dark scalar fields can condensate into the states of high occupation numbers.
We have found the influence of $\langle S(x) \rangle$ to the changes of the Higgs properties at the tree level.
At the scales $M\sim M_{Pl}$ the direct searches detection of the SDM  is constrained ($\sim 10^{-61}$) to be well below the experimental sensitivity today as well as for nearest future. However, the effect of new physics would be visible if $M\sim O$ (100 TeV) that would be an effective test for the future  colliders.
The confrontation of these bounds with other models and the experimental observables will constitute the further test to the SDM-Higgs model.

Although the work has tried to shed the light in the physics of the FBS, still much need to be understood, especially the interactions between the scalars and the gravitational attracted forces, before we can make the predictions for the observables and the consequences related to these observables. 
For appropriate values of the free parameters and the experimental data in (\ref{e41}) and (\ref{e402}), the model considered in this paper may open the wide and precise cosmological phenomenology, providing the stability of the BS  and the influence of the  $\langle S(x) \rangle$ on the latter. The (thermal) photon spectrum and the momentum distribution of the leptons would probe directly the density, the size and the mass scales of the BS.
We found that a significance of the SDM contribution to the FBS compared to that of the Higgs boson is very small, $\sim O(10^{-34})$. However, because of the SDM life-time much bigger than $\tau_h$, the origin of $\nu$'s registration at the Earth laboratory is the decay of the long-lived SDM via the DP's decaying to these neutrinos. The BS as the source of the neutrinos is located in our Galaxy at the distances $\leq$ 1 kpc. 
The additional knowledge may be required to understand the mechanism of the FBS just after the born of the Universe. 
If the scalar boson stars exist, then their cosmological consequences and the relations with particle physics should be investigated. To that end, one has to use the general principles of quantum field theory flowing to the effective field theory to understand the behaviour of the stars in the evolution from the early stage of their formation up today. 
Although the results are presented only for the scalar bosons stars, the mechanism described in this paper should also be applied to a much broader class of massive giant cosmological objects, such as the Fermi stars, the Neutron stars, the Proto-neutron stars which also exhibit dynamical scale symmetry.

\end{document}